\documentclass{elsart5p}
\usepackage{graphicx}
\usepackage{amssymb}

\begin{document}

\begin{frontmatter}



\title{Remnants of compact binary mergers}


%\author[hd]{W. Domainko} and M. Ruffert}

\author[hd]{W. Domainko \corauthref{cor1}}
\author[ed]{and M. Ruffert}
\corauth[cor1]{wilfried.domainko@mpi-hd.mpg.de}
\address[hd]{Max-Planck Institute for Nuclear Physics, Saupfercheckweg 1, 69117 Heidelberg, Germany}
\address[ed]{School of Mathematics, University of
Edinburgh, Edinburgh EH9 3JZ, Scotland, UK}

\address{}

\begin{abstract}
We investigate the long-term evolution and observability of remnants originating
from the merger of compact binary systems and discuss the differences to
supernova remnants. Compact binary mergers expel much smaller amounts of mass
at much higher velocities, as compared to supernovae, which will affect the
dynamical evolution of their remnants. The ejecta of mergers consist of very 
neutron rich nuclei. Some of these neutron rich nuclei will produce 
observational signatures in form of gamma ray lines during their decay. The
composition of the ejecta might even give interesting constraints about the
internal structure of the neutron star. We 
further discuss the possibility that merger remnants appear as recently 
discovered 'dark accelerators' which are extended TeV sources which lack emission in other bands.

\end{abstract}


\end{frontmatter}

\section{Introduction}
\label{}

Compact binary mergers are ideal targets to study matter in
extreme circumstances. Remnants left behind by such cosmological explosions
are promising sites to investigate merger ejecta and its interaction with 
a surrounding medium. As remnants survive for a considerable time \newline ($\sim
10^6$ years until they are diluted by turbulence of the surrounding ISM) 
it is possible to explore them long after the corresponding merger event
and one should be able
to find them as relics from nearby mergers which can
be investigated in greater detail.

Observations have revealed that compact binaries exist 
in the universe.
Several neutron star - neutron 
star (NSNS)
binaries are identified (Stairs 2004 \cite{stairs04}) and it is expected that they will 
spiral in as a result of gravitational wave emission
and merge subsequently. Also possible signatures of the 
coalescence of compact objects have been observed. NSNS 
mergers were proposed as central engines of cosmological
GRBs already about two decades ago (e.g. Paczy\'nski 1986 \cite{paczynski86}, Eichler et al. 1989 \cite{eichler89})
and are nowadays
considered as the most promising explanation for the subclass of short and
hard bursts (Lee et al. 2005 \cite{lee05}). Theoretical progress on the physics of NSNS  and neutron star black hole (NSBH) merger was made with extensive numerical simulations (e.g. Ruffert 1996 \cite{ruffert96}, 1997 \cite{ruffert97}, Lee \& Klu\'zniak 1999 \cite{lee99} and Rosswog 2005 \cite{rosswog05}). These simulations revealed that compact binary mergers eject a small fraction of their mass (10$^{-4}$ - 0.1 M$_{\odot}$) with a velocity of a significant fraction of the speed of light. The evolution of the ejecta on timescales of the first few days was investigated by Li \& Paczy\'nski (1998) \cite{li98}.

Gamma ray burst remnants (GRBRs) were studied in the past due to their aspherical shape (Ayal \& Piran 2001 \cite{ayal01}) and their ability to create giant HI holes in the interstellar medium (Efremov et al. 1998 \cite{efremov98}, Loeb \& Perna 1998 \cite{loeb98}).   
Very recently extended objects radiating in very high energy gamma rays without significant
emission in other bands which were discovered by the H.E.S.S. collaboration (Aharonian et al. 2005a \cite{aharonian05a}, 2006 \cite{aharonian06}), were suggested to be
GRBRs (Domainko \& Ruffert 2005 \cite{domainko05}, Atoyan et al. 2006 \cite{atoyan06}).

\section{Dynamical Evolution}

Very energetic events exploding into a surrounding medium will leave remnants 
(e.g. Chevalier 1977 \cite{chevalier77}). 
In general remnants of cosmological explosions expand freely until the mass of the displaced ambient medium equals the mass of the ejecta. After this has happened the remnants thermalise and expand in the Sedov phase.
Since compact binary mergers also have an outflow component in form of jets the evolution of the remnants will start highly aspherical. However, this outflow component will become quasi-spherical on timescales of less than one year (Livio \& Waxman 2000 \cite{livio00}). 
Mergers eject a much smaller amount of material at a much
larger velocity as compared to supernova explosions. Consequently the dynamical evolution of a merger remnant will differ from the evolution of a supernova remnant (SNR).
Due to the small mass of ejecta in mergers the free
expansion phase of a merger remnant will be short. Therefore it will enter
the Sedov phase much earlier and with a much higher velocity than a SNR.
We show this with an example.  We calculate the expansion of a merger remnant and a SNR in a hot and thin medium with a density of n = 10$^{-2}$
cm$^{-3}$ and
a temperature of T = $10^7$ K (the sound velocity in the adopted surrounding medium is 479 km/s). The choice of the embedding medium is motivated by the recent observations of short GRBs exploding in the intra-cluster medium of galaxy clusters (Gehrels et al. 2005 \cite{gehrels05}, Bloom et al. 2006 \cite{bloom06}, Prochaska et al. 2006 \cite{prochaska06}). For a merger with  a mass ejected of m$_{ej}$ = $5 \times 10^{-3}$ 
M$_{\odot}$ and an ejection velocity of
v$_{ej}$ = $1.5 \times 10^{5}$ km/s (parameters are chosen to correspond to a
kinetic energy of $10^{51}$ erg) the free expansion phase will only last for
about 12 years. In the early Sedov phase,
the remnant still expands with a high 
Mach-number of 40 after 100 years (see Fig. 1.). The situation is different for a
supernova exploding into the same environment. For an extensive discussion of SNR in hot, thin media see Dorfi \& V\"olk (1996) \cite{dorfi96} and Tang \& Wang (2005) \cite{tang05}. A supernova type Ia 
with m$_{ej}$ = 1.4 M$_{\odot}$ and v$_{ej}$ = $5 \times 10^{3}$ km/s 
(E$_{kin}$ again 10$^{51}$ erg) will feature a free expansion phase of 
about 1360
years. The Mach-number of the corresponding remnant at the beginning of the
Sedov phase is about 9 (see again Fig. 1.). As one interesting result from 
these considerations, we find that
merger remnants expand with high Mach-numbers also in hot, thin environments which makes
them potential sites for particle acceleration even in media with large
sound velocities.

\begin{figure}[ht]\label{mach}
\includegraphics[width=6.0cm,angle=-90]{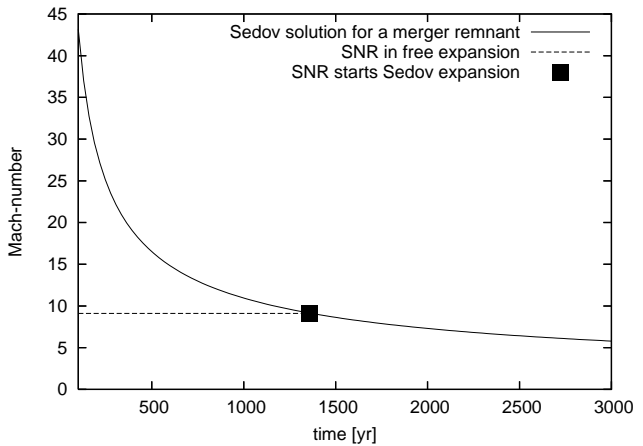}
\caption{Comparison between the evolution of a compact binary
merger remnant and a supernova
remnant (SNR). The merger remnant starts the Sedov phase much earlier with much
higher Mach-number. After the SNR has entered the Sedov expansion (at an age of
1360 years), the dynamical
evolution of both remnants is very similar since it only depends on the
properties of the embedding medium and the 
mechanical energy of the explosion (here adopted to be the same for both cases).
The major difference between the expansion of SNRs and merger remnants is in the time interval before the SNR reaches the Sedov phase (time $<$ 1360 years). The curve of the temporal evolution of the Mach number in SNRs and merger remnants coincide in this plot after 1360 years. 
For more details see main text.}
\end{figure}

\section{The ejecta}

\subsection{Gamma ray line emission}

The ejecta of compact binary mergers consist of exceptionally neutron rich
material. This might result in the production and distribution of heavy
r-process elements (Lattimer \& Schramm 1974 \cite{lattimer74}, Freiburghaus et al. 1999 \cite{freiburghaus99}). Some of these heavy nuclei will be radioactive and may
emit observational signatures in form of gamma ray lines during their decay (Clayton \& Craddock 1965 \cite{clayton65}, Qian et al. 1999 \cite{qian99}, Domainko \& Ruffert 2005 \cite{domainko05}). 
Of special interest for the case of merger remnants are nuclei which have similar half life times as the average merger rate in the Galaxy, since these nuclei feature most likely the strongest emission in galactic remnants. The average merger rate in the Galaxy was estimated to about one event every \newline (0.5 - 7)$\times$10$^4$ years using the properties of the known galactic NSNS systems (Kalogera et al. 2004 \cite{kalogera04}). This result is also in rough agreement with the constraints from the observations of short GRBs (Guetta \& Piran 2006 \cite{guetta06}). Hence in an optimistic scenario we expect one to several merger remnants with an age of $\gtrsim 10^4$ years in the galaxy. 
To estimate the expected signal of gamma ray lines which results from the decay of r-process nuclei we use the model of Qian et al. (1999 \cite{qian99}).
In Table 1 we give the line strength of gamma ray lines connected to the decay of some heavy nuclei with half life times exceeding 1000 years. The values are calculated for the advantageous case of an age of the remnant of 5000 years, a distance of 2 kpc and an initial mass of each individual nuclei of 10$^{-5}$ M$_{\odot}$. 
The yields of actinides in mergers may be significantly larger than in core collapse supernovae since for low values of the relative electron number density of the ejecta these events will mainly expel very heavy r-process nuclei (Ruffert et al. 1997 \cite{ruffert97}). The previously discussed 
gamma ray line sources could be promising targets for future instruments using focusing optics for soft gamma rays with Laue lenses or Fresnel lenses (Kn\"odlseder 2006 \cite{knoedelseder06}, Skinner 2001 \cite{skinner01}).

\subsection{Origin of the ejecta}

The amount of ejecta depends on system parameters like the mass and the
spin of the two components (Janka et al. 1999 \cite{janka99}, Rosswog 2005 \cite{rosswog05}). One result of this effect will be that the origin 
of the ejecta inside the
neutron star is different for different systems. For systems featuring
a higher mass loss, part of the ejecta will be expelled from deeper layers 
inside the
neutron star as compared to systems which feature a smaller mass loss.   
Nucleosynthesis of material of different origin inside the neutron star leads to a different chemical composition (Goriely et al. 2005 \cite{goriely05}). Hence the
composition of the ejecta can give interesting information about the internal 
structure of neutron stars. Even material from the core region of the neutron star might be expelled. In NSBH binaries the neutron star could be shredded due to several close encounters with the black hole until it reaches its minimal mass in timescales of tens of seconds (Davies et al. 2005 \cite{davies05}). Neutron stars with minimal mass may explode (Blinnikov et al. 1984 \cite{blinnikov84}) which will appear as electromagnetic transients in the keV range with luminosities of up to 10$^{47}$ erg/s and a typical duration of around tens of seconds (Sumiyoshi et al. 1998 \cite{sumiyoshi98}). Such events are indeed observed after some short GRBs (e.g. GRB 050724 Barthelmy et all. 2005 \cite{barthelmy05}, GRB 050911 Page et al. 2006 \cite{page06}). In those systems the GRB could be launched by the first accretion event on the black hole (see also Davies et al. 2005 \cite{davies05} and Page et al. 2006 \cite{page06}) and the flare on timescales of 10 seconds thereafter might be interpreted as the explosion of the neutron star with minimal mass. In the later event material from the innermost regions of the original neutron star would be ejected into the surrounding (Sumiyoshi et al. 1998 \cite{sumiyoshi98}).

\begin{table}
\centering

\caption{Properties of gamma ray lines}
\vspace{0.5cm}
\begin{tabular}{lcccccc}

\hline
r-process &   &   $\tau$   &  &  E$_{\gamma}$   &  &  F$_{\gamma}$   \\
nucleus   &  & [10$^3$ yr] &  & [keV] &  &  [10$^{-7}$ $\gamma$ cm$^{-2}$ s$^{-1}$]   \\
\hline

$^{226}$Ra & & 2.31 & & 609 & & 0.78 \\
$^{229}$Th & & 10.6 & & 40.0 & & 0.61 \\
$^{251}$Cf & & 1.30 & & 177  & & 0.09 \\
\hline

\end{tabular}
\end{table}
\vspace{0.5cm}

\section{High energy gamma ray signatures}

A common feature of long as well as short bursts is a highly relativistic outflow in form of jets. Hence processes related to relativistic shocks will accelerate particles in both system in a comparable way which can then lead to very high energy radiation in their remnants.  Differences in the appearance of GRB remnants will result from the fact that these two systems differ considerably in the nature and energetics of the central engine, the mass of ejecta and the environment into which they are evolving.

\subsection{Energetics of the relativistic shocks}

Relativistic shocks which are the connecting feature of both types of GRBs cannot accelerate particles to a spectrum represented by a single power law but will cause a spectrum with a break below some particular energy. Accelerated electrons (synchrotron) cool very quickly after the shock becomes non relativistic. During the evolution of the remnant, energy dependent diffusion will steepen the spectrum of the remaining cosmic ray protons and inject these high energy particles into the ISM. Interactions of the cosmic ray protons with the target material of the ISM through inelastic proton-proton collisions and subsequent $\pi^0$ decay ($\pi^0$ channel) will then lead to an extended, center filled TeV source with suppressed synchrotron emission.  The total energy in cosmic rays which are accelerated by the relativistic jets of long GRBs and then injected into the ambient medium through the above mentioned scenario, can exceed the entire energy in accelerated particles generated by the non relativistic shocks of average SNRs by a factor of $\sim$100. Thus remnants of such events will very likely represent strong TeV sources. For a comprehensive description of the before mentioned model see Atoyan et al. (2006 \cite{atoyan06}). 
In contrast to typical long bursts the energy of short bursts can vary significantly. Several short events are found to be less powerful than typical long bursts by an order of magnitude (Prochaska et al. 2006 \cite{prochaska06}) whereas others are comparable in energetics to long events (e.g. GRB 051221A Soderberg et al. 2006 \cite{soderberg06}). Consequently the energy in cosmic rays available for production of very high energy gamma rays will in many cases be considerably smaller in remnants of short burst than in remnants of typical long bursts.

\subsection{Signature of the ejecta}

Important for distinguishing remnants from long and short bursts is the fact that this two distinct classes of bursts are launched by remarkably different events: long bursts are related to the death of very massive stars (Galama et al. 1998 \cite{galama98}) and the nowadays favoured model for short bursts is connected to the mergers of compact binary stars (e.g. Lee et al. 2005 \cite{lee05}). These diverse central engines will distribute a significantly unequal amount of ejecta.
Hypernovae of very massive stars will expel a fair fraction of their mass into the explosion site. Mergers, as discussed in the introduction, will eject an amount of mass which is orders of magnitude smaller than in an average supernova. Therefore observational signatures of the ejecta in other wavebands in merger remnants will be much weaker than in hypernova remnants.

\subsection{Impact of the environment}

A major difference between the TeV gamma ray production in remnants of long and short GRBs is the environment in which those events explode. The environment is important in particular  for the production of TeV gamma rays through the $\pi^0$ channel since it acts as the target material for the inelastic proton-proton collision. Long bursts on the one hand are generally connected to very massive stars so they will happen within or close to star forming regions. These locations are favorable for the hadronic $\pi^0$ channel since high density target material in form of molecular clouds is often present there. For compact binaries, on the other hand, it takes a much longer time until they finally merge and it is expected that these events happen in various environments, even in the low densities surrounding the galactic plane. Hence merger remnants will only produce noteworthy TeV emission in the case of location in reasonably dense media. 

\subsection{Summary}

To summarize the above discussion, remnants of both type of GRBs may look like extended, center-filled TeV emitters with suppressed fluxes in the radio to GeV wavebands (Atoyan et al. 2006 \cite{atoyan06}). In the case of compact binary mergers the very high energy gamma ray emission will be less luminous but potential signatures of the ejecta in other wavebands will be much weaker as compared to the case of hypernovae.

For possible identifications of merger remnants with particular very high energy gamma ray sources it will be indispensable to identify the location of the sources in the galaxy. HESS J1303-631 is presumably connected to the Cen OB 1 association of very young stars (Aharonian et al. 2005b \cite{aharonian05b}) and since compact binary mergers are likely related to an older stellar population it is not a promising candidate for a merger remnant.

\section*{Acknowledgements}

We acknowledge the support of the European Commission through grant number
RII3-CT-2003-506079 (HPC-Europa). We want to thank 
Marialuisa Aliotta, Roland Diehl, Thomas Janka, Ewald M\"uller, Alex Murphy, Sabine Schindler 
and Philip Woods for enlightening discussions. 

% \appendix

% \section{}
% \label{}

% Bibliographic references with the natbib package:
% Parenthetical: \citep{Bai92} produces (Bailyn 1992).
% Textual: \citet{Bai95} produces Bailyn et al. (1995).
% An affix and part of a reference:
%   \citep[e.g.][Ch. 2]{Bar76}
%   produces (e.g. Barnes et al. 1976, Ch. 2).


\begin{thebibliography}{}

% \bibitem[Names(Year)]{label} or \bibitem[Names(Year)Long names]{label}.
% (\harvarditem{Name}{Year}{label} is also supported.)

\bibitem{aharonian05a} Aharonian, F., Akhperjanian, A. G., Aye, K.-M., et al. A New Population of Very High Energy Gamma-Ray Sources in the Milky Way. Science 307, 1938-1942, 2005a.

\bibitem{aharonian05b} Aharonian, F., Akhperjanian, A. G., Aye, K.-M., et al. Serendipitous discovery of the unidentified extended TeV $\gamma$-ray source HESS J1303-631. A\&A, 439, 1013-1021, 2005b.

\bibitem{aharonian06} Aharonian, F., Akhperjanian, A. G., Bazer-Bachi, A. R., et al. The H.E.S.S. Survey of the Inner Galaxy in Very High Energy Gamma Rays. ApJ, 636, 777-797, 2006.

\bibitem{atoyan06} Atoyan, A., Buckley, J. \& Krawczynski, H. A Gamma-Ray Burst Remnant in Our Galaxy: HESS J1303-631. ApJ, 642, L153-L156, 2006.

\bibitem{ayal01} Ayal, S. \& Piran, T. Remnants from Gamma-Ray Bursts. ApJ, 555, 23-30, 2001.

\bibitem{barthelmy05} Barthelmy, S. D., Chincarini, G., Burrows, D. N., et al. An origin for short $\gamma$-ray bursts unassociated with current star formation. Natur, 438, 994-996, 2005.

\bibitem{blinnikov84} Blinnikov, S. I., Novikov, I. D., Perevodchikova, T. V., Polnarev, A. G. Exploding Neutron Stars in Close Binaries. SvAL, 10, 177-179, 1984.	


\bibitem{bloom06} Bloom, J. S., Prochaska, J. X., Pooley, D., et al. Closing in on a Short-Hard Burst Progenitor: Constraints from Early-Time Optical Imaging and Spectroscopy of a Possible Host Galaxy of GRB 050509b. ApJ, 638, 354-368, 2006.

\bibitem{chevalier77} Chevalier, R. A. The interaction of supernovae with the interstellar medium. ARA\&A, 15, 175-196, 1977.

\bibitem{clayton65} Clayton, D. D. \& Craddock, W. L. Radioactivity in Supernova Remnants. ApJ, 142, 189-200, 1965.

\bibitem{davies05} Davies, M. B., Levan, A. J. \& King, A. R. The ultimate outcome of black hole-neutron star mergers. MNRAS, 356, 54-58, 2005.

\bibitem{domainko05} Domainko, W. \& Ruffert, M. Long-term remnant evolution of compact binary mergers. A\&A, 444, L33-L36, 2005.

\bibitem{dorfi96} Dorfi, E. A., V\"olk, H. J. Supernova remnant dynamics and particle acceleration in elliptical galaxies. A\&A, 307, 715-725, 1996.

\bibitem{efremov98} Efremov, Y. N., Elmgreen, B. G., \& Hodge, P. W. Giant Shells and Stellar Arcs as Relics of Gamma-Ray Burst Explosions. ApJ, 501, L163-L165, 1998.

\bibitem{eichler89} Eichler, D., Livio, M., Piran, T. \& Schramm, D. N. Nucleosynthesis, neutrino bursts and gamma-rays from coalescing neutron stars. Natur, 340, 126-128, 1989.

\bibitem{freiburghaus99} Freiburghaus, C., Rosswog, S., \& Thielemann, F.-K. R-Process in Neutron Star Mergers. ApJ, 525, L121-L124, 1999.

\bibitem{galama98} Galama, T. J., Vreeswijk, P. M., van Paradijs, J. et al. An unusual supernova in the error box of the gamma-ray burst of 25 April 1998. Natur, 395, 670-672, 1998.

\bibitem{gehrels05} Gehrels, N., Sarazin, C. L., O'Brien, P. T., et al. A short $\gamma$-ray burst apparently associated with an elliptical galaxy at redshift z = 0.225. Natur, 437, 851-854, 2005.

\bibitem{goriely05} Goriely, S., Demetriou, P., Janka, H.-Th., Pearson, J. M. \& Samyn, M. The r-process nucleosynthesis: a continued challenge for nuclear physics and astrophysics. NuPhA, 758, 587-594, 2005.

\bibitem{guetta06} Guetta, D. \& Piran, T. The BATSE-Swift luminosity and redshift distributions of short-duration GRBs. A\&A, 453, 823-828, 2006.

\bibitem{janka99} Janka, H.-T., Eberl, T., Ruffert, M., Fryer, C. L. Black Hole-Neutron Star Mergers as Central Engines of Gamma-Ray Bursts. ApJ, 527, L39-L42, 1999.

\bibitem{kalogera04} Kalogera, V., Kim, C., Lorimer, D. R., et al. The Cosmic Coalescence Rates for Double Neutron Star Binaries. ApJ, 601, L179-L182, 2004. erratum, ApJ, 614, L137-L138, 2004.

\bibitem{knoedelseder06} Kn\"odlseder, J. GRI: the gamma-ray imager mission. Proceedings of the SPIE, Volume 6266, p. 61, astro-ph/0608149, 2006.

\bibitem{lattimer74}
Lattimer, J. M. \& Schramm, D. N. Black-hole-neutron-star collisions. ApJ, 192, L145-L147, 1974.

\bibitem{lee99} Lee, W. H. \& Klu\'zniak, W. L. Newtonian Hydrodynamics of the Coalescence of Black Holes with Neutron Stars. I. Tidally Locked Binaries with a Stiff Equation of State. ApJ, 526, 178-199, 1999.

\bibitem{lee05} Lee, W. H., Ramirez-Ruiz, E. \& Granot, J. A. Compact Binary Merger Model for the Short, Hard GRB 050509b. ApJ, 630, L165-L168, 2005.

\bibitem{li98} Li, L.-X. \& Paczy\'nski, B. Transient Events from Neutron Star Mergers. ApJ, 507, L59-L62, 1998.

\bibitem{livio00} Livio, M. \& Waxman, E. Toward a Model for the Progenitors of Gamma-Ray Bursts. ApJ, 538, 187-191, 2000.

\bibitem{loeb98} Loeb, A. \& Perna, R. Are H i Supershells the Remnants of Gamma-Ray Bursts? ApJ, 503, L35-L37, 1998.

\bibitem{paczynski86} Paczy\'nski, B. Gamma-ray bursters at cosmological distances. ApJ, 308, L43-L46, 1986.

\bibitem{page06} Page, K. L., King, A. R., Levan, A. J., et al. GRB 050911: A Black Hole-Neutron Star Merger or a Naked GRB. ApJ, 637, L13-L16, 2006. 

\bibitem{prochaska06} Prochaska, J. X., Bloom, J. S., Chen, H.-W., et al. The Galaxy Hosts and Large-Scale Environments of Short-Hard Gamma-Ray Bursts. ApJ 642, 989-994, 2006.

\bibitem{qian99} Qian, Y.-Z., Vogel, P. \& Wasserburg, G. J. Probing r-Process Production of Nuclei Beyond $^{209}$BI with Gamma Rays. ApJ, 524, 213-219, 1999.

\bibitem{soderberg06} Soderberg, A. M., Berger, E., Kasliwal, M., et.al. The Afterglow, Energetics, and Host Galaxy of the Short-Hard Gamma-Ray Burst 051221a. ApJ 650, 261-271, 2006.

\bibitem{rosswog05} Rosswog, S. Mergers of Neutron Star-Black Hole Binaries with Small Mass Ratios: Nucleosynthesis, Gamma-Ray Bursts, and Electromagnetic Transients. ApJ, 634, 1202-1213, 2005.

\bibitem{ruffert96} Ruffert, M., Janka, H.-T. \& Schaefer, G. Coalescing neutron stars - a step towards physical models. I. Hydrodynamic evolution and gravitational-wave emission. A\&A, 311, 532-566, 1996.

\bibitem{ruffert97} Ruffert, M.; Janka, H.-T.; Takahashi, K.; Schaefer, G. Coalescing neutron stars - a step towards physical models. II. Neutrino emission, neutron tori, and gamma-ray bursts. A\&A, 319, 122-153, 1997.

\bibitem{skinner01} Skinner, G. K. Diffractive/refractive optics for high energy astronomy. I. Gamma-ray phase Fresnel lenses. A\&A, 375, 691-700, 2001. 

\bibitem{stairs04} Stairs, I. H. Pulsars in Binary Systems: Probing Binary Stellar Evolution and General Relativity. Science, 304, 547-552, 2004.

\bibitem{sumiyoshi98} Sumiyoshi, K., Yamada, S., Suzuki, H. \& Hillebrandt, W. The fate of a neutron star just below the minimum mass: does it explode? A\&A, 334, 159-168, 1998.

\bibitem{tang05} Tang, S. \& Wang, Q. D. Supernova Blast Waves in Low-Density Hot Media: A Mechanism for Spatially Distributed Heating. ApJ, 628, 205-209, 2005.

\end{thebibliography}
\end{document}